\documentclass[journal=nanoletters,manuscript=article]{achemso}

\usepackage{chemformula} 
\usepackage[T1]{fontenc} 
\usepackage{siunitx}

\author{Carlos Errando-Herranz} \altaffiliation{Equal contribution} \email{carloseh@kth.se} 
\author{Eva Sch\"oll} \email{escholl@kth.se} \altaffiliation{Equal contribution}
\affiliation[KTH Royal Institute of Technology] {Quantum Nanophotonics, KTH Royal Institute of Technology, Roslagstullsbacken 21, 10691 Stockholm, Sweden}
\author{Rapha\"{e}l Picard}
\author{Micaela Laini} 
\affiliation[Heriot-Watt University] {Institute for Photonics and Quantum Sciences, SUPA, Heriot-Watt University, Edinburgh EH14 4AS, UK}
\author{Samuel Gyger} 
\author{Ali W. Elshaari}
\author{Art Branny}
\author{Ulrika Wennberg}
\author{Sebastien Barbat}
\author{Thibaut Renaud} 
\affiliation[KTH Royal Institute of Technology] {Quantum Nanophotonics, KTH Royal Institute of Technology, Roslagstullsbacken 21, 10691 Stockholm, Sweden}
\author{Mauro Brotons-Gisbert}
\author{Cristian Bonato}
\affiliation[Heriot-Watt University] {Institute for Photonics and Quantum Sciences, SUPA, Heriot-Watt University, Edinburgh EH14 4AS, UK}
\author{Brian D. Gerardot}
\affiliation[Heriot-Watt University] {Institute for Photonics and Quantum Sciences, SUPA, Heriot-Watt University, Edinburgh EH14 4AS, UK}
\author{Val Zwiller}
\author{Klaus D. J\"ons}
\affiliation[KTH Royal Institute of Technology] {Quantum Nanophotonics, KTH Royal Institute of Technology, Roslagstullsbacken 21, 10691 Stockholm, Sweden} \email{klausj@kth.se}

\title[Resonance fluorescence from waveguide--coupled strain--localized two-dimensional quantum emitters]
  {Resonance fluorescence from waveguide--coupled strain--localized two--dimensional quantum emitters}

\begin{document}
\begin{abstract}
Efficient on--chip integration of single--photon emitters imposes a major bottleneck for applications of photonic integrated circuits in quantum technologies. 
Resonantly excited solid--state emitters are emerging as near--optimal quantum light sources, if not for the lack of scalability of current devices. 
Current integration approaches rely on cost--inefficient individual emitter placement in photonic integrated circuits, rendering applications impossible. 
A promising scalable platform is based on two--dimensional (2D) semiconductors. 
However, resonant excitation and single--photon emission of waveguide--coupled 2D emitters have proven to be elusive. 
Here, we show a scalable approach using a silicon nitride photonic waveguide to simultaneously strain--localize single--photon emitters from a tungsten diselenide (\ch{WSe2}) monolayer and to couple them into a waveguide mode.
We demonstrate the guiding of single photons in the photonic circuit by measuring second--order autocorrelation of g$^{(2)}(0)=0.150\pm0.093$ and perform on--chip resonant excitation yielding a g$^{(2)}(0)=0.377\pm0.081$.
Our results are an important step to enable coherent control of quantum states and multiplexing of high--quality single photons in a scalable photonic quantum circuit.

\end{abstract}

\section{Introduction}
Large--scale on--chip quantum technologies are crucial to realize the long--standing goals of photonic quantum information processing, such as quantum communication~\cite{borregaard_one-way_2019}, quantum simulation~\cite{aspuru-guzik_photonic_2012}, and quantum computing based on cluster state generation~\cite{Rudolph_why_2017, ladd_quantum_2010}.
A promising route towards large--scale quantum information processing relies on single--photon qubits, and is based on quantum emitters, memories, and detectors interconnected via photonic integrated circuits (PICs)~\cite{flamini_photonic_2018}.

Single--photon emitter integration into PICs has been achieved by embedding quantum dots into III--V PIC platforms~\cite{hepp_semiconductor_2019}, with limited scalability due to their optical loss, large waveguide bend radius, and low fabrication yields. 
To utilize the scaling offered by PICs, pick--and--place techniques have been developed to integrate III--V semiconductor quantum dots~\cite{kim_hybrid_2020} and diamond color centers~\cite{mouradian_scalable_2015} into silicon (Si) and silicon nitride (SiN) waveguide platforms. 
A drawback of this approach lies on the stringent requirements for emitter fabrication and precise pick--and--place of individual emitters, which drastically limit the scalability of this technology.

A promising candidate to overcome the current scalability limitations of quantum PICs is based on two--dimensional (2D) materials~\cite{stanford_emerging_2018}. 
In particular, transition metal dichalcogenides (TMDs)~\cite{tonndorf_single-photon_2015,kumar_strain-induced_2015,srivastava_optically_2015,he_single_2015,koperski_single_2015,chakraborty_voltage-controlled_2015} enable hundreds of single--photon emitters by a single pick--and--place transfer using localized strain~\cite{branny_deterministic_2017,palacios-berraquero_large-scale_2017}.
Efforts towards 2D TMD single--photon emitter integration into PICs are on the rise, including transfer of a tungsten diselenide (\ch{WSe2}) monolayer on the facet of a titanium--indiffused lithium niobate waveguide with a large mode size~\cite{white_atomically-thin_2019}, and on top of a lossy plasmonic slot waveguide~\cite{blauth_coupling_2018}.
More scalable approaches have been initiated, such as coupling of single--photons from a $90$~nm thick gallium selenide layered semiconductor in \ch{SiN} waveguides~\cite{tonndorf_-chip_2017}, photoluminescence from a \ch{WSe2} monolayer into a \ch{SiN} waveguide~\cite{peyskens_integration_2019}, and emission from hexagonal boron nitride (hBN) in an aluminum nitride waveguide~\cite{kim_integrated_nodate}.
However, single--photon emission into a photonic circuit from deterministic strain--localized quantum emitters has proven to be elusive, let alone resonant excitation of 2D quantum emitters through a PIC, a prerequisite for future initialization and coherent control of quantum states~\cite{warren_coherent_1993} and for the generation of highly indistinguishable photons~\cite{Somaschi_near-optimal_2016, ding_-demand_2016}.

Here, we overcome these challenges by (i) inducing strain--localized quantum emitters at the waveguide edges, (ii) multiplexing emitters into the same waveguide mode, (iii) demonstrating waveguide--coupled single--photon emission, and (iv) performing resonant excitation of a single quantum emitter through the waveguide.
Our results show the potential of combining 2D semiconductors with PICs towards large--scale quantum technologies, by realizing crucial building blocks for future complex circuits.

\section{Results}
\subsection{Strain--localized emitters in a 2D semiconductor}

\begin{figure}[t]
	\centering
	\includegraphics[width=1\columnwidth]{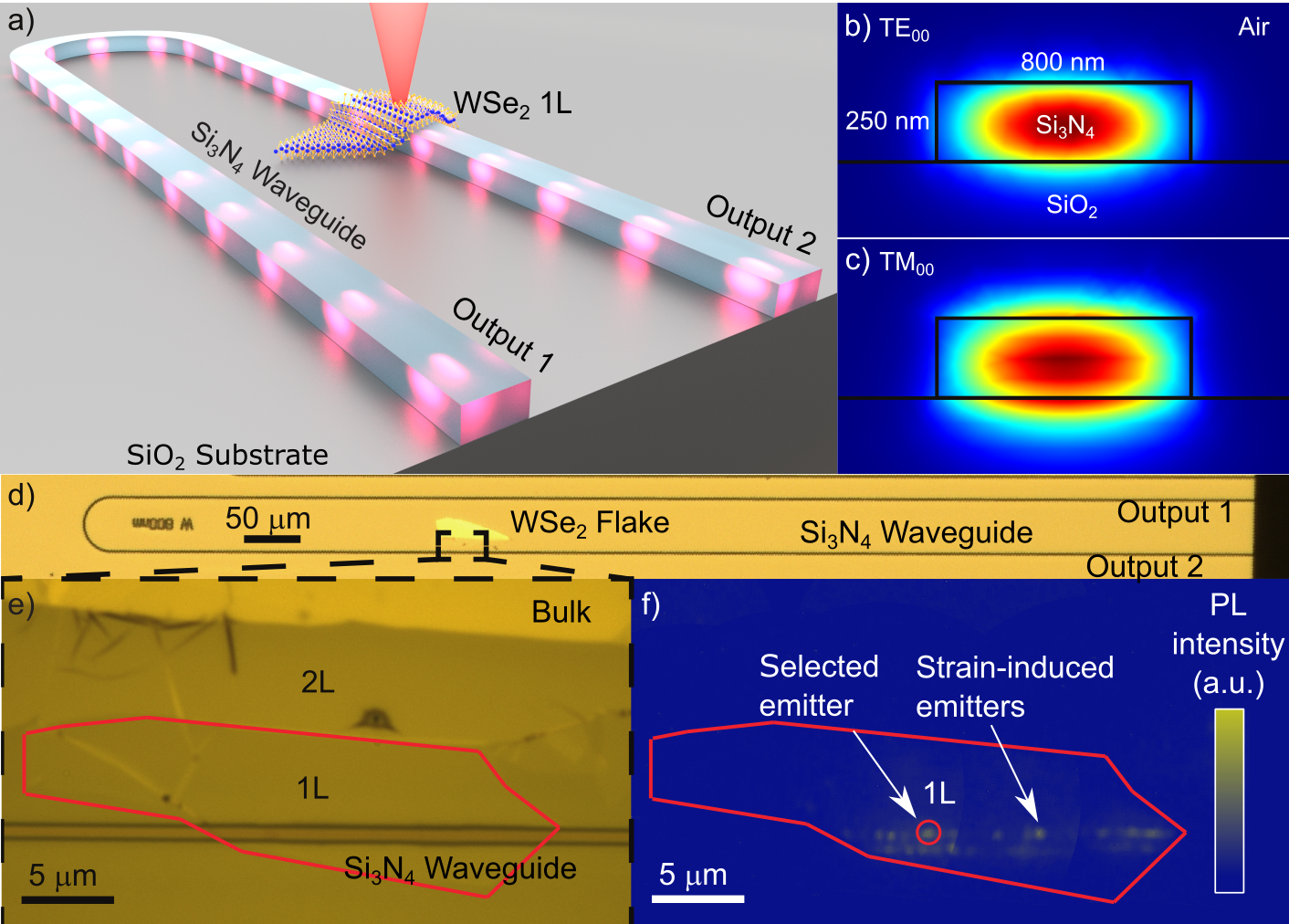}
	\caption{a) Artistic illustration of the coupled \ch{WSe2} monolayer (1L) single--photon emitter and the \ch{Si3N4} waveguide. b) Finite element method eigenmode simulation of the fundamental quasi--TE and c) quasi--TM waveguide modes at FIXME~nm wavelength. d) Microscope image of the \ch{Si3N4} waveguide with e) zoom in of the \ch{WSe2} flake. The monolayer is marked in red (1L). f) Photoluminescence with de--focused excitation shows strain--localized emitters along the waveguide edges. The emitter used for further experiments is marked with a red circle.}
	\label{fig:sample}
\end{figure}

Figure~\ref{fig:sample}a shows a schematic of our sample, consisting of a U--shaped \ch{Si3N4} waveguide on a \ch{SiO2} bottom cladding. 
The designed waveguide geometry supports the fundamental quasi--TE and quasi--TM waveguide modes as shown in Fig.~\ref{fig:sample}b \& c. 
The microscope image in Fig.~\ref{fig:sample}d gives an overview of the whole structure, with cleaved facets and an exfoliated \ch{WSe2} monolayer (1L) placed on top of the waveguide using a dry--transfer method~\cite{Castellanos-Gomez_deterministic_2014} (see Fig.~\ref{fig:sample}e and Supplementary).
Figure~\ref{fig:sample}f shows photoluminescence from emitters in the sample under de--focused excitation, recorded using a CCD camera with a $700$~nm long pass filter to remove backscattered laser light. 
The measurements were performed with a modular setup consisting of a closed--cycle cryostat at~$6$~K where the sample was placed on a piezoelectric movable stage, a spectrometer, and a Hanbury Brown and Twiss (HBT) second--order correlation measurement setup, as shown in Fig.~\ref{fig:setup}a. 
A detailed description of the setup is given in the Supplementary.
The emitters in the monolayer were excited from the top through a microscope objective with a red pulsed laser ($638$~nm) with variable repetition rate of $5$-$80$~MHz.
In line with reported strain--localization of single--photon emitters ~\cite{branny_deterministic_2017, palacios-berraquero_large-scale_2017}, we observe two lines of spatially localized emitters along the waveguide edges. 
Since the flake remains continuous across the waveguide and shows no signs of rupturing or wrinkling, we suspect that waveguide roughness along the edges is responsible for creating local strain gradients that localize the emission. This way we are able to multiplex several emitters into the same waveguide mode by a single transfer step (see Supplementary for more waveguide--coupled emitters). 

\begin{figure}[tbh]
	\centering
	\includegraphics[width=1\columnwidth]{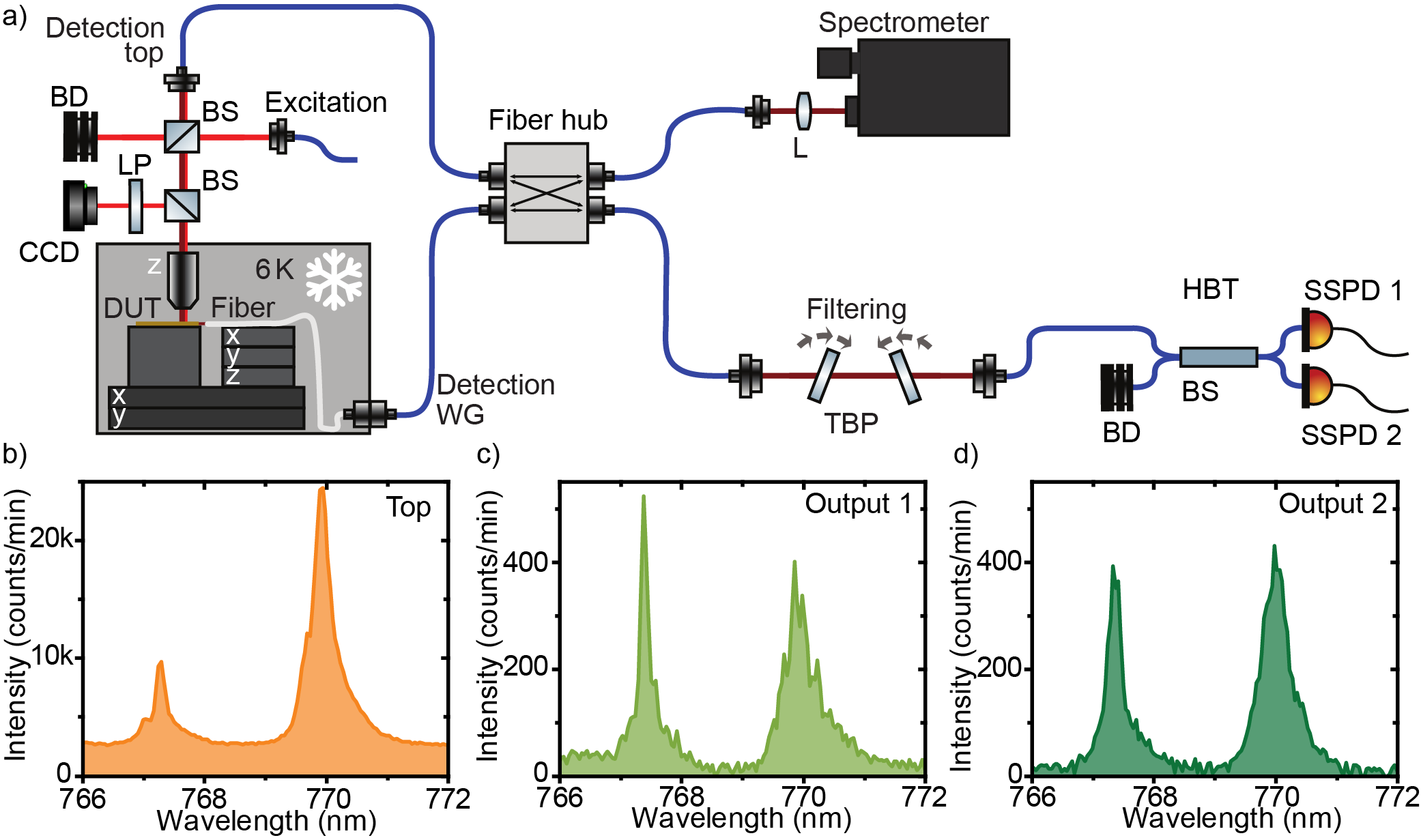}
	\caption{a) Modular setup consisting of a red laser excitation, a confocal detection path (detection top) and a second detection path from the waveguide facet through a lensed fiber (detection WG). In the fiber hub, the signals can be routed to the spectrometer or the Hanbury Brown and Twiss setup (HBT), which includes a free--space filtering by two tunable bandpass filters (TBP). DUT device under test; BS beam splitter; L lens; BD beam dump; SSPD superconducting single--photon detector. b) Spectra from emitter 1 taken from top, c) through the waveguide from output 1, and d) from output 2.}
	\label{fig:setup}
\end{figure}

By focusing the excitation laser onto the sample and using a confocal microscopy setup, we recorded the photoluminescence spectra of single emitters.
Figure~\ref{fig:setup}b shows the spectrum of an emitter marked in Fig.~\ref{fig:sample}f collected out of plane of the waveguide through the objective (detection top). 
To identify the peaks, we performed polarization resolved photoluminescence spectroscopy (see Supplementary), which indicates that both lines most likely stem from different emitters.
Figure~\ref{fig:setup}c and d show the spectra at the same location collected through the two waveguide output ports. 
The line at $770~\si{\nm}$, emitter 1, is used for all further measurements under non-resonant excitation. 

A common signature of a two--level system is saturation of the emission intensity with increasing excitation power, shown in Fig.~\ref{fig:g2}a in a double--logarithmic plot. 
Fitting the data as described in the Supplementary, we extracted a saturation power of $414\pm48$~\si{\nano\W}. 
All further measurements were performed with an excitation power of $1.4$~\textmu W, located at the start of the saturation plateau for a best trade--off between high emission intensity and increasing background.

\subsection{Single--photon emission from a 2D emitter}
\begin{figure}[tbh]
	\centering
	\includegraphics[width=1\columnwidth]{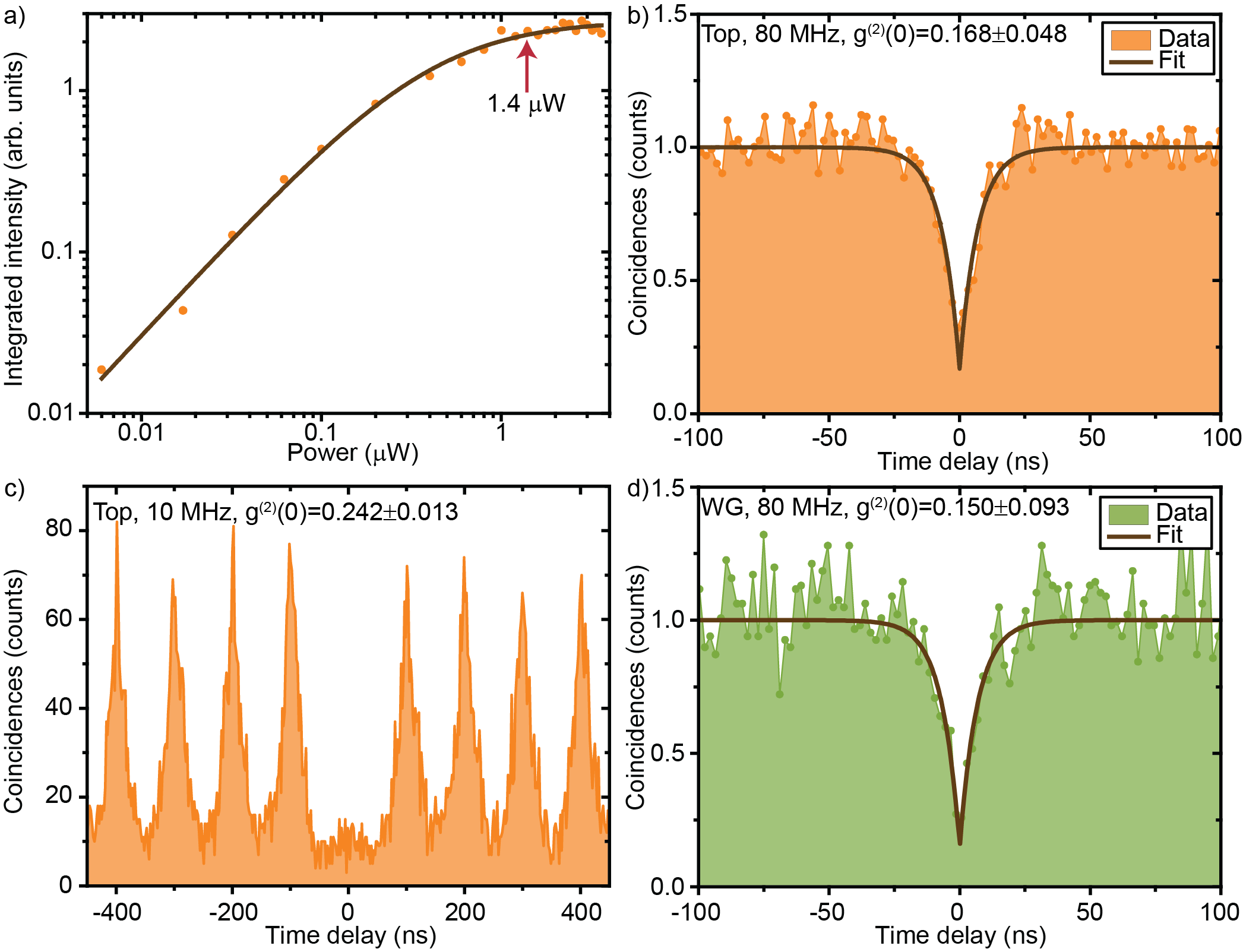}
	\caption{a) Power series for emitter 1 with repetition rate $80$~MHz. For all correlation measurements, the emitter was excited with $1.4$~\textmu W, i.e. at the start of the saturation plateau. b) Second--order autocorrelation measurement from top, c) with a lower repetition rate ($10$~MHz), and d) through the waveguide output 1.}
	\label{fig:g2}
\end{figure}

To confirm single--photon emission, we performed a second--order autocorrelation measurement on the emitted signal of the line at $770~\si{\nm}$, filtered by two overlapping tunable bandpass filters (bandwidth $20~\si{\nm}$), and with a time binning of $2048$~ps.
Although the emitters were excited with a $80$~\si{MHz} repetition rate pulsed laser, our second--order autocorrelation measurement, shown in Fig.~\ref{fig:g2}b, resembles a measurement under a continuous--wave laser excitation.
We investigated this by measuring the emission lifetime with a lower laser repetition rate of $5$~\si{MHz} (see Supplementary). 
Fitting the data with a double--exponential decay, we extracted a lifetime of $18.3\pm1$~\si{\ns}, which is significantly longer than the separation of two consecutive excitation pulses of $12.5$~\si{\ns} corresponding to a repetition rate of $80$~\si{\MHz}. 
This in turns leads to a strong overlap between neighboring peaks in the histogram, which can not be distinguished from the noise on the Poisson level. Our simulation results (see Supplementary) suggest that under this circumstance, the pulsed second--order autocorrelation measurement can be treated like a continuous--wave measurement.
Fitting the data with the formula given in the Supplementary yields a g$^{(2)}(0)$ of $0.168\pm0.048$, well below $0.5$ (see Fig.~\ref{fig:g2}b), which demonstrates the single--photon nature of the light emission from our 2D emitter. Additionally, we measured the second--order autocorrelation with a repetition rate of $10$~\si{\MHz} shown in Fig.~\ref{fig:g2}c, yielding a g$^{(2)}(0)$ of $0.242\pm0.013$ without post--selection (See Supplementary for the analysis).

\subsection{Single--photons from a 2D emitter through a \ch{Si3N4} waveguide}
Next, we investigated waveguide coupling of single--photon emission from 2D \ch{WSe2} emitters. 
We simulate the coupling efficiency from the emitter, approximated by a planar dipole, into the waveguide modes (see Supplementary). 
By varying dipole orientation and positions along the top edge of the waveguide we calculated its emission into the fundamental quasi--TE (TE00) and quasi--TM (TM00) waveguide modes. 
The unidirectional coupling efficiency to the fundamental modes when the dipole is located at the edge of the waveguide is, on average for all possible in--plane dipole orientations, $0.32\,\%$ and $0.34\,\%$ to the TE00 and TM00 mode, respectively. 
Experimentally, we collect the waveguide--coupled emission using a lensed single--mode fiber mounted on an adjacent, independently movable piezoelectric stage, and aligned to one of the waveguide ends.
For all waveguide coupled measurements, the fiber was coupled to output 1, marked in Fig.~\ref{fig:sample}a.
We performed a second--order autocorrelation measurement through the waveguide, shown in Fig.~\ref{fig:g2}d yielding g$^{(2)}(0)=0.150\pm0.093$.
This value shows no degradation with respect to the free--space g$^{(2)}(0)$ value, and demonstrates strain--localized single--photon emission into a waveguide.

\subsection{Resonance fluorescence of waveguide--coupled 2D quantum emitters} 

\begin{figure}
	\centering
	\includegraphics[width=1\columnwidth]{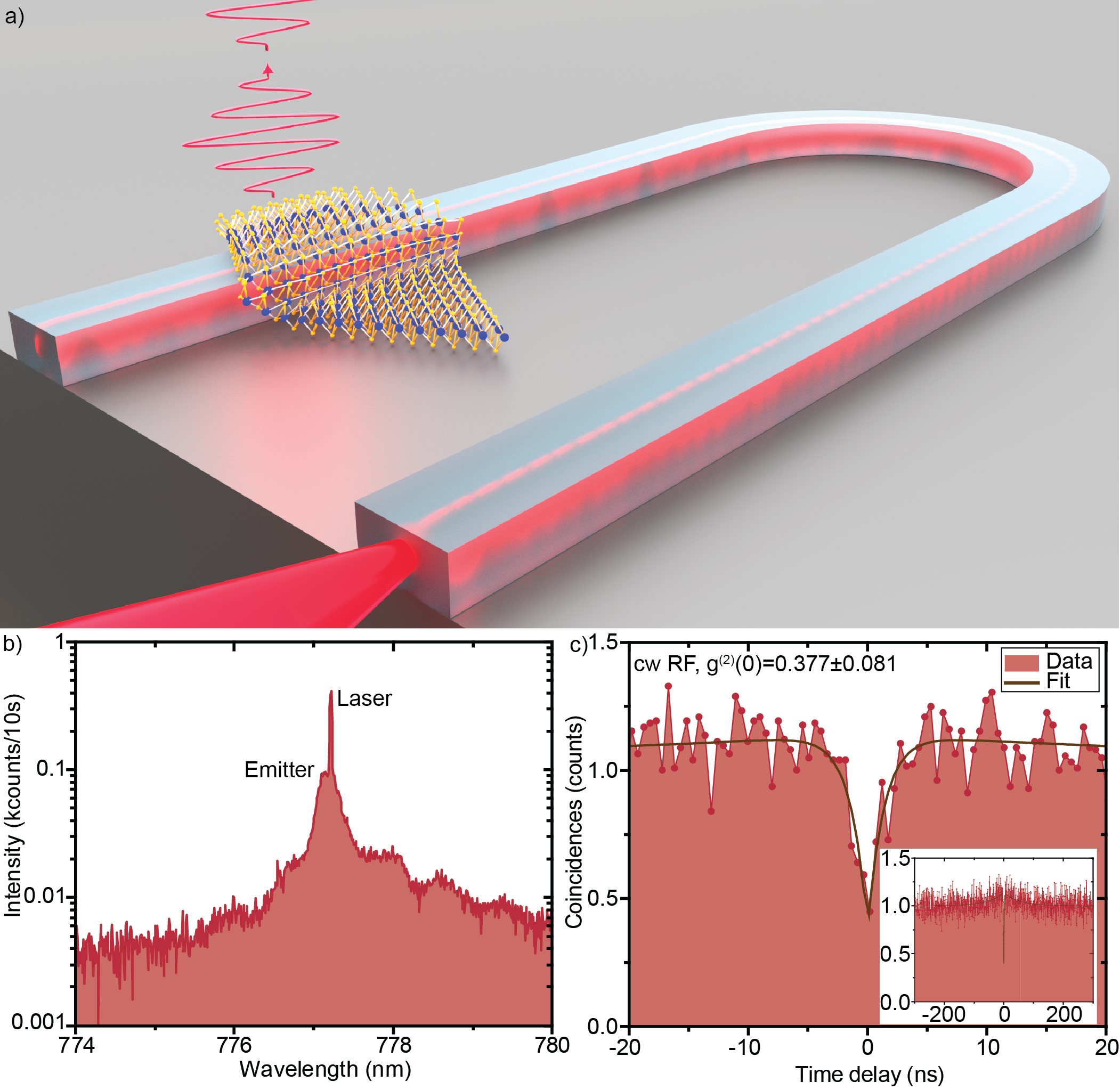}
	\caption{a) Artistic illustration of the coupled \ch{WSe2} monolayer single--photon emitter on the \ch{Si3N4} waveguide. The 2D emitter is excited with a continuous--wave (cw) laser coupled to the waveguide. The emitted signal is detected from top through a microscope objective. b) Resonance fluorescence (RF) spectrum of emitter 2 and residual laser in a semilogarithmic plot. c) Second--order autocorrelation measurement under resonant excitation through the waveguide and detection from top showing clear single--photon emission. Inset: Same measurement for a longer time window showing bunching originating from spectral diffusion.}
	\label{fig:RF}
\end{figure}

Finally, we used our integrated device to perform resonant excitation using side--excitation~\cite{Muller_resonance_2007} through the waveguide output\,1. 
So far only off--chip confocal resonant excitation of \ch{WSe2} and hBN emitters have been reported, requiring data post-processing by either post-selection of time intervals when the emitter was on resonance to combat spectral wandering~\cite{kumar_resonant_2016} or laser background subtraction~\cite{Konthasinghe_rabi_2019}. 
Another approach has been to spectrally filter the zero-phonon line together with the resonant laser and only collecting the phonon sideband~\cite{tran_resonant_2018}.

Here, we achieve sufficient laser suppression in our waveguide--coupled circuit to measure the second--order correlation function without the need of background subtraction nor complex post--processing analysis. 
Instead, we perform on-the-fly optimization of polarization suppression, and only stop and restart the measurement for realignment if the on-the-fly suppression malfunctions.
Our resonant excitation and detection scheme of a waveguide--coupled 2D quantum emitter is artistically illustrated in Fig.~\ref{fig:RF}a. 
A continuous--wave diode laser with a linewidth of $50$\,kHz is coupled via a lensed fiber into the waveguide which guides the excitation light to the monolayer. 
The emitted signal is collected from top by a microscope objective. 
In this scheme, a large portion of the laser remains in the waveguide and only the light scattered by the waveguide surface is collected by the objective.
Further spatial suppression of laser light is achieved by fiber coupling the collected signal.
To distinguish the resonance fluorescence signal from the remaining spatially overlapping scattered laser light, the laser is suppressed in a polarization suppression setup (see Supplementary for a detailed description). 
The resonance fluorescence of emitter 2 as well as the remaining laser light is shown in Fig.~\ref{fig:RF}b, with deliberately non--optimal laser suppression for visualization.  
This shows the slight mismatch between the laser wavelength and the emitter spectrum, which stems from a spectral shift of the emitter when the laser is on resonance.
We then performed a second--order autocorrelation measurement with a time binning of $512$~ps (see Supplementary for more details), shown in Fig.~\ref{fig:RF}c, yielding a g$^{(2)}=0.377\pm0.081$, indicating clear single--photon emission from the emitter under pure resonant excitation. 
The non-zero value at zero time delay stems mainly from remaining laser scattering. 
Furthermore, the emitter shows light bunching on the timescale of $50$~ns (inset of Fig.~\ref{fig:RF}c), originating from spectral diffusion.

\section{Discussion}

\begin{figure}
	\centering
	\includegraphics[width=1\columnwidth]{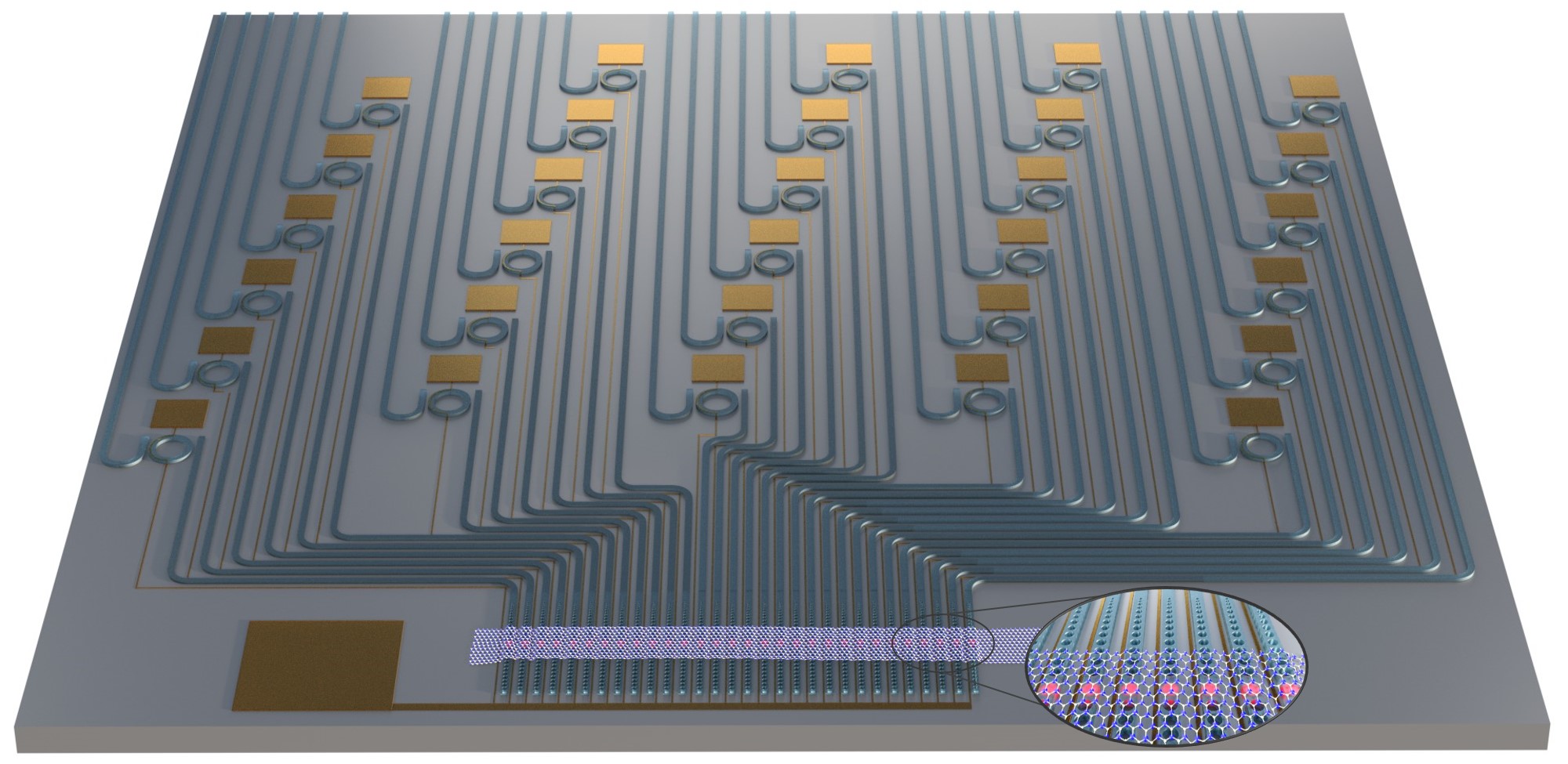}
	\caption{Vision for a large--scale on--chip 2D emitter based quantum light source. An array of emitters from a single TMD monolayer transfer are strain--localized and coupled into the waveguides via on--chip cavities. Reconfigurable ring resonators filter the indistinguishable single photons, which are then ready to be coupled off--chip for quantum communications, or remain on--chip for cluster--state quantum computing or simulation. Inset: zoom--in to the TMD monolayer covering several on--chip photonic cavities.}
	\label{fig:vision}
\end{figure}

Our single--photon emitter localization using local strain gradients enables scalable integration of quantum emitters in PICs.
However, in the current design, the emitters are located at the edge of the waveguide, yielding sub--optimal coupling efficiency to the fundamental waveguide modes (see simulations in the Supplementary). 
More efficient coupling can be achieved by localizing the emitter centered at the top of the waveguide, with an average directional coupling efficiency of 2.48\,\% and 3.01\,\% in the TE00 and TM00 modes (see Supplementary), or by encapsulation of the emitter, with up to 22\,\%  directional coupling for TE00.
The compatibility between the localization scheme and the optimal PIC geometry demands non-trivial solutions, which currently stand as remaining challenges hindering efficient coupling. 
A solution may involve inducing emitters with a helium focused ion beam~\cite{klein_site-selectively_2019}, or point-localizing emitters using the strain arising from pillars, gaps and terminations along waveguides. 
Alternative methods might be the use of cavities, dielectric screening~\cite{raja_coulomb_2017}, or Moir\'e trapped excitons~\cite{tran_evidence_2019,seyler_signatures_2019,brotons-gisbert_spin-layer_2019, baek_highly_2020}.
Therefore a path towards quantum PIC can take many forms where a single TMD monolayer generates many emitters, overcoming the current bottlenecks of single emitter pick-and-place methods.
Figure~\ref{fig:vision} shows our vision for a large--scale quantum light source based on the presented technology. 
A single transfer of a TMD monolayer creates an array of localized emitters efficiently coupled via on--chip cavities into waveguides, where the indistinguishable single photons travel and get filtered by reconfigurable ring resonators.
The output is a large number of indistinguishable single--photons, which can be routed into the optical fiber network for quantum communications, or remain on--chip for cluster--state quantum computing or simulation.

The limited single--photon characteristics under resonant excitation stem mainly from remaining scattered laser which can not be suppressed in the polarization suppression setup. 
Due to the current setup and device, the polarization of the input laser as well as the scattered laser from the waveguide can not be well controlled and therefore not perfectly suppressed. 
These challenges can be overcome by using polarization maintaining fibers or free--space coupling of the laser to the waveguide, on--chip polarization control, and/or reduction of laser scattering by fabricating waveguides with smoother sidewalls.
Next steps towards coherent control of PIC coupled 2D quantum emitters will require on--chip resonance fluorescence by excitation and detection through the waveguide, measuring the two--photon interference visibility from such emitters, and pulsed resonant excitation of these quantum emitters. 

In addition to multiplexing, large-scale quantum photonic circuits require two--photon interference between photons emitted from independent on--chip sources. 
However, fluctuations in the electrostatic and strain environment of individual single--photon emitters make emitted photons spectrally differ, hampering quantum interference.
To address this issue, spectral tuning is required, and has been demonstrated using the strain induced by piezoelectric~\cite{iff_strain-tunable_2019,kim_position_2019} and capacitive~\cite{grosso_tunable_2017} actuators, or by electric field tuning using the Stark effect~\cite{noh_stark_2018}.

\section{Conclusion} 
Quantum photonic integrated circuits provide a scalable and cost--efficient route to increasingly complex quantum systems, and constitute an enabling platform for applications such as quantum key distribution, quantum simulation, and cluster--state quantum computing.
We have developed a hybrid deterministic integration method of single--photon emitters in 2D materials into silicon-based photonic circuits by exploiting the creation of strain--localized quantum emitters at the edges of a photonic waveguide. 
Our proof--of--principle structure maintains a single--photon purity of $0.150\pm0.093$, and resonance fluorescence with g$^{(2)}=0.377\pm0.081$.
These experimental results and proposed designs provide a promising hybrid integration platform with promising scaling prospects, crucial for large--scale quantum integrated circuits.


\begin{acknowledgement}
This project has received funding from the European Union's Horizon 2020 research and innovation program under grant agreement No. 820423 (S2QUIP), EPSRC (EP/P029892/1) and ERC (No. 725920).
C.E. acknowledges funding from the Swedish Research Council (2019-00684).
K.D.J. acknowledges funding from the Swedish Research Council (VR) via the starting Grant HyQRep (Ref 2018-04812) and The Göran Gustafsson Foundation (SweTeQ).
B.D.G. is supported by a Wolfson Merit Award  from  the  Royal  Society  and  a  Chair  in  Emerging Technology from the Royal Academy of Engineering.
C.E. is currently with the Quantum Photonics Laboratory, Massachusetts Institute of Technology.
\end{acknowledgement}

\section {Author contribution} 
C.E. and E.S. contributed equally to this work.
B.D.G. and K.D.J. conceived the experiments. 
C.E., S.G., M.L., and K.D.J. simulated and designed the device.
C.E., S.G., and A.W.E. developed and carried out the waveguide fabrication and characterization.
A.B., M.L., R.P., S.B., and T.R. developed and carried out the monolayer transfer.
K.D.J and E.S. built the cryogenic optics setup.
E.S., R.P., U.W., and K.D.J. carried out the optical measurements.
E.S. and K.D.J. analyzed the data. 
C.E., E.S., A.B, and K.D.J. wrote the manuscript, with help from all the authors.
B.D.G. and K.D.J. supervised the project.

\section{Competing interests}
The authors declare no competing financial interests.

\begin{suppinfo}
The following files are available free of charge.
\begin{itemize}
  \item Sample geometry and fabrication, experimental setup for top excitation, polarization resolved photoluminescence spectroscopy, waveguide coupling simulations, multiplexed emitters in the waveguide, lifetime measurement, power--dependent photoluminescence measurements, analysis of second--order autocorrelation measurements under non-resonant excitation, methods for second--order autocorrelation measurements under resonant excitation.
\end{itemize}

\section{Sample geometry and fabrication}
The waveguide length is $2.2$~mm, with a bend radius of $25$~\textmu m, and a cross-section (height $\times$ width) of $250 \times 800$~nm.
This waveguide cross-section supports 2 quasi-TE and 2 quasi-TM modes, its fundamental quasi-TE mode index being $\text{n}_\text{TE00}=1.72$.
The flake is positioned at a $750$~\textmu m distance from one of the waveguide ends, and covers a waveguide section about $20$~\textmu m long.

The sample fabrication started with a $250$~nm thin stoichiometric \ch{Si3N4} film on $3.3$~\textmu m \ch{SiO2} on a silicon substrate (Rogue Valley Microdevices).
The waveguides were fabricated using electron beam lithography followed by \ch{CHF3}-based reactive ion etching, resist stripping, and sample cleaving.
Flux zone grown \ch{WSe2} crystals (from 2D semiconductors) were then exfoliated, and monolayers were identified under a microscope and transferred using a polydimethylsiloxane (PDMS) dry stamp process~\cite{Castellanos-Gomez_deterministic_2014}.

\section{Experimental setup for top excitation}
The setup is shown in Fig.~$2$a in the main text.
For all measurements, the sample was placed inside a low--vibration closed--cycle cryostat on piezoelectric xy--positioners and cooled to $6~\si{\kelvin}$. 
For excitation, a red ($638~\si{\nano\meter}$) pulsed laser diode with variable repetition rate of $5-80~\si{\MHz}$ was used, which was focused onto the sample using a $50\times, NA=0.81$ microscope objective. 
The photoluminescence of the excited emitter was detected in two ways: from the top using free--space optics or through the waveguide. 
The part of the photoluminescence emitted to the top was collected through the same microscope objective and then coupled into a fiber. 
The part coupled to the \ch{Si3N4} waveguide was coupled into a lensed fiber (OZOptics, 780HP, working distance $13\pm1~\si{\micro\meter}$) positioned near the cleaved facet of one of the waveguide ends. 
The lensed fiber, which was mounted on an individual xyz--positioner stack, was pre--aligned to the waveguide by sending a narrow linewidth laser at $770~\si{\nano\meter}$ through the fiber and maximizing the signal at the other output of the waveguide with the CCD camera through the microscope objective. 
The fine alignment was done by maximizing the emitter signal on the spectrometer.
The fiber--coupled signal was either sent to the CCD of a spectrometer (grating $600$~lines/mm) or into a fiber--based Hanbury Brown and Twiss (HBT) type setup to measure the second--order autocorrelation function. 
This setup consists of a $50:50$ fiber beamsplitter connected to two superconducting single--photon detectors (Single Quantum) with efficiencies of $50$\%, $60$\%, timing jitters of $20$ and $30$~ps, and dark count rates of $0.006$ and $0.017~\text{cts}\slash\text{s}$. 
For all correlation measurements, a single line was filtered from the spectrum using two overlapping free--space tunable bandpass filters with a bandwidth of $20~\si{\nm}$. 
For excitation with a green laser ($532~\si{\nano\meter}$), the excitation was coupled into the setup using a dichroic longpass mirror with the edge at $695~\si{\nano\meter}$ (not shown in the setup).
To investigate the localization of the emitters in the \ch{WSe2} monolayer, we excited the sample with a de--focused laser (lens with $f=300~\si{\milli\meter}$, not shown). 
The emitted signal was sent onto a CCD camera using a flip--in beamsplitter, and the accompanying backscattered excitation laser was filtered with a $700~\si{\nano\meter}$ long pass filter.

\section{Polarization resolved photoluminescence spectroscopy}

\begin{figure}
	\centering
	\includegraphics[width=1\columnwidth]{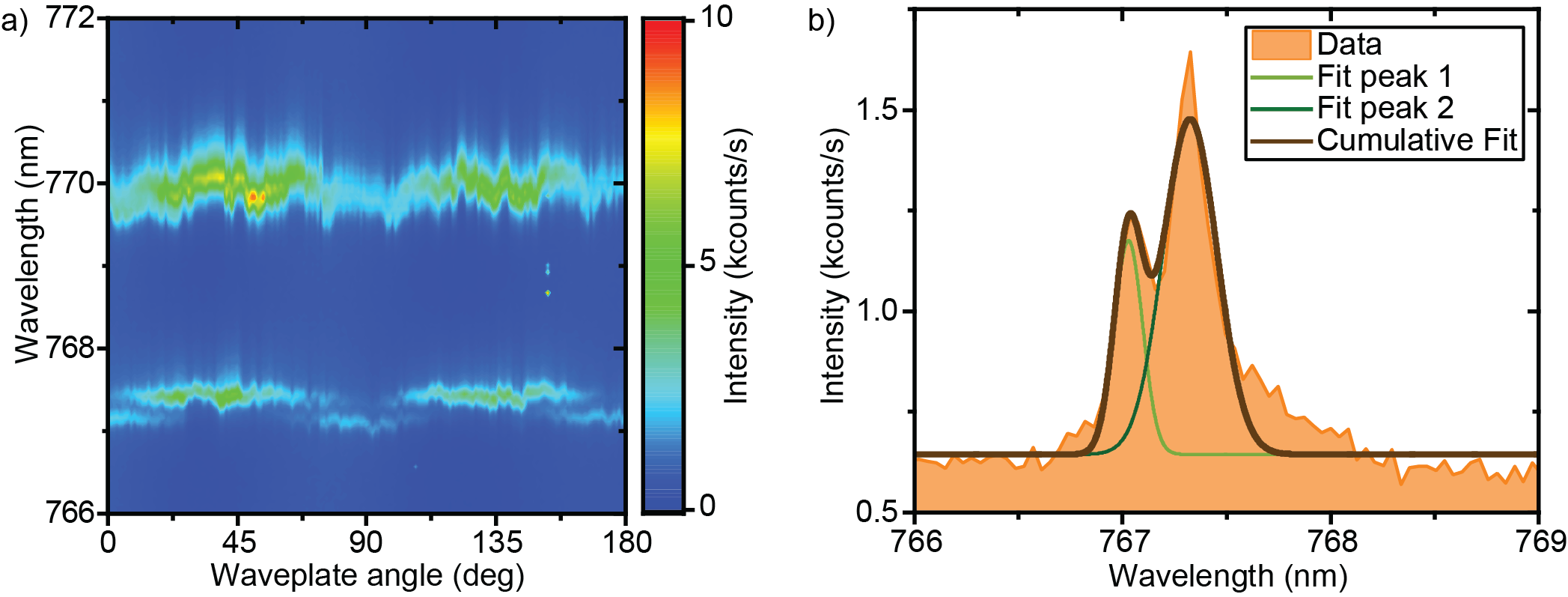}
	\caption{a) Map of emitter 1 under changing half--wave plate angle. b) Spectrum of the line at 767 nm and a halfwave plate angle of $70\si{\degree}$. A fit reveals a fine--structure splitting of $624\pm31\si{\micro\eV}$.}
	\label{suppfig:Pol}
\end{figure}

To identify the peaks of the spectra in Fig.\,2b-d in the main manuscript, we measured the polarization resolved photoluminescence. 
We placed a halfwave plate and a fixed linear polarizer in front of the spectrometer and recorded the spectra while automatically rotating the wave plate, as shown in Fig.\,\ref{suppfig:Pol}a. 
Only the line at $767\,\si{\nm}$ is showing a clear fine--structure splitting. Figure\,\ref{suppfig:Pol}b shows the spectrum of this line at a halfwave plate angle of $70\si{\degree}$, where both components are well visible. 
Fitting this data with two Gaussians, reveals a fine--structure splitting of $624\pm31\si{\micro\eV}$, indicating an exciton. 
The line at $770\,\si{\nm}$ shows intensity fluctuations but not clear fine structure splitting, which indicates that this line still belongs to an excitonic state where the intensity of one fine--structure component is too dim to see.
This leads us to believe that those two lines stem from separate emitters.

\section{Waveguide coupling simulations}

\begin{figure}
	\centering
	\includegraphics[width=0.98\columnwidth]{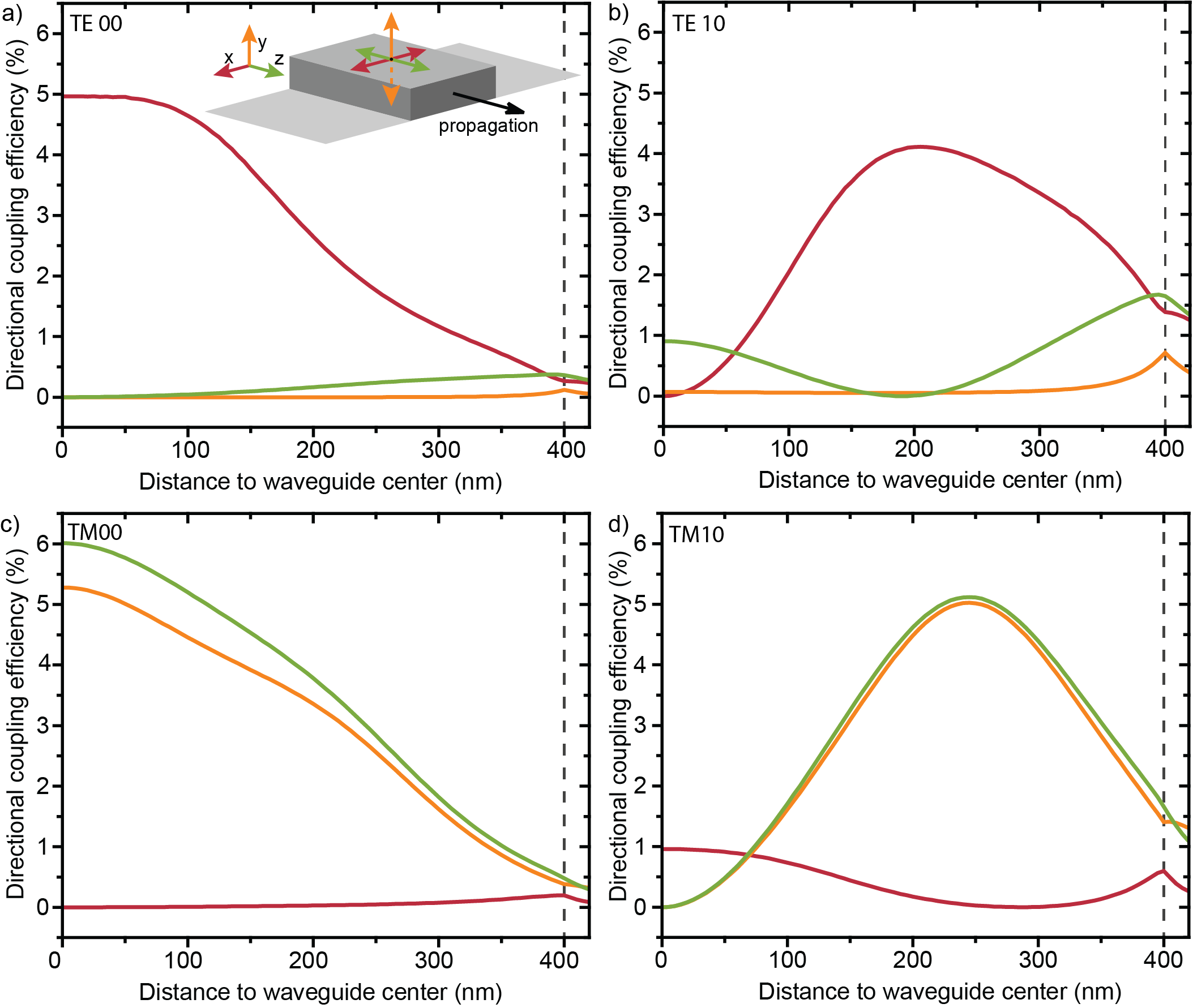}
	\caption{FDTD Simulation of the coupling efficiency to the guided modes (TE00, TE10, TM00, TM10) of the three different dipole orientations depending on the location on the waveguide. The dipole is located $5$~nm above the waveguide. $x=0$~nm is at the center of the waveguide.}
	\label{suppfig:coupling}
\end{figure}

We simulated the fabricated structure using a 3D-FDTD solver (Lumerical), and modeled the emitter as a dipole $5$~nm above the $800$~nm wide waveguide. 
The position of the dipole is swept across the width of the waveguide, and the directional coupling efficiency is shown in Fig.\,\ref{suppfig:coupling}, where $x=0$~nm represents the center of the waveguide, and $x=400$~nm the edge.
We swept the dipole across half of the waveguide width since the system is symmetric, and negative displacements will yield the same coupling conditions (i.e. one can mirror our results along the $x=0$~nm line to obtain the full waveguide width sweep).

We average over the two dipole orientations parallel to the substrate (x and z in Fig.~\ref{suppfig:coupling}a). 
The directional coupling emission with the emitter at the edge of the waveguide is $0.3\%$ ($3.3\%$) to the TE00 (all) modes. 
The maximum simulated directional coupling efficiency into the TE00 (all) mode for averaged dipole orientation in--plane (x and z in Fig.~\ref{suppfig:coupling}a) was $2.5\%$ ($7.9\%$).  

\section{Multiplexed emitters in the waveguide}

\begin{figure}
	\centering
	\includegraphics[width=0.96\columnwidth]{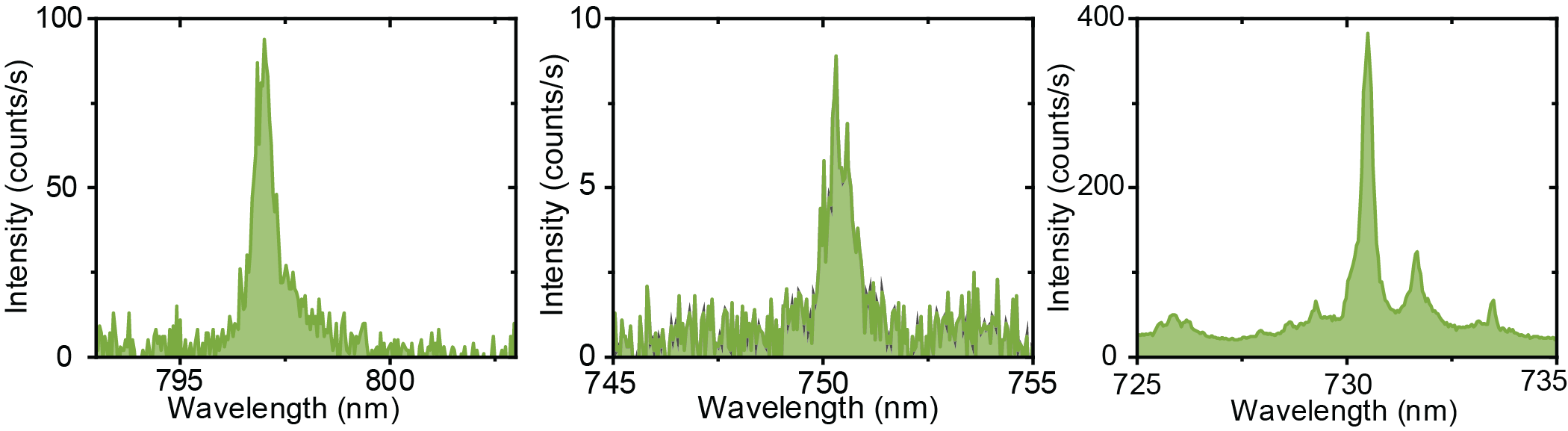}
	\caption{Spectra of 3 waveguide--coupled emitters from the same 2D flake, PL measured through the waveguide. These emitters are in addition to the 2 emitters studied in the main text. }
	\label{suppfig:mux}
\end{figure}

Figure\,\ref{suppfig:mux} shows waveguide--coupled emission from three different 2D emitters, in addition to the two emitters in the main text, strain--localized from a single monolayer transfer. 

\section{Lifetime measurement}

\begin{figure}
	\centering
	\includegraphics[width=0.5\columnwidth]{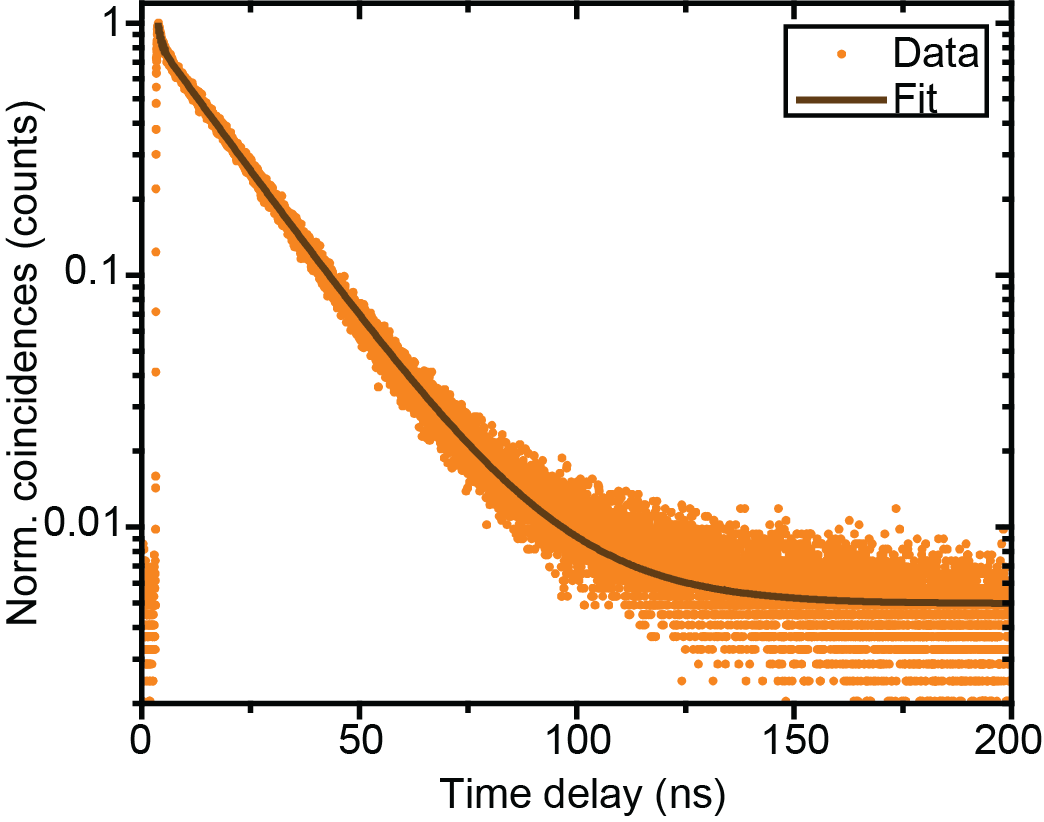}
	\caption{Lifetime measurement of emitter 1 under red $5~\si{\MHz}$ excitation and fitted using a double--exponential curve.}
	\label{suppfig:lifetime}
\end{figure}

To understand why our pulsed second--order correlation measurement looks like a cw measurement, we measured the lifetime of the excited state of emitter 1 with a lower repetition rate of $5~\si{\MHz}$ compared to the second--order correlation measurement, and correlated the detected signal with the trigger from the laser in a standard time--correlated single--photon counting experiment. 
Figure\,\ref{suppfig:lifetime} shows the data together with an double--exponential fit yielding to a lifetime of $18.3\pm1$~\si{\ns}.

\section{Power--dependent photoluminescence measurements}

\begin{figure}
	\centering
	\includegraphics[width=0.5\columnwidth]{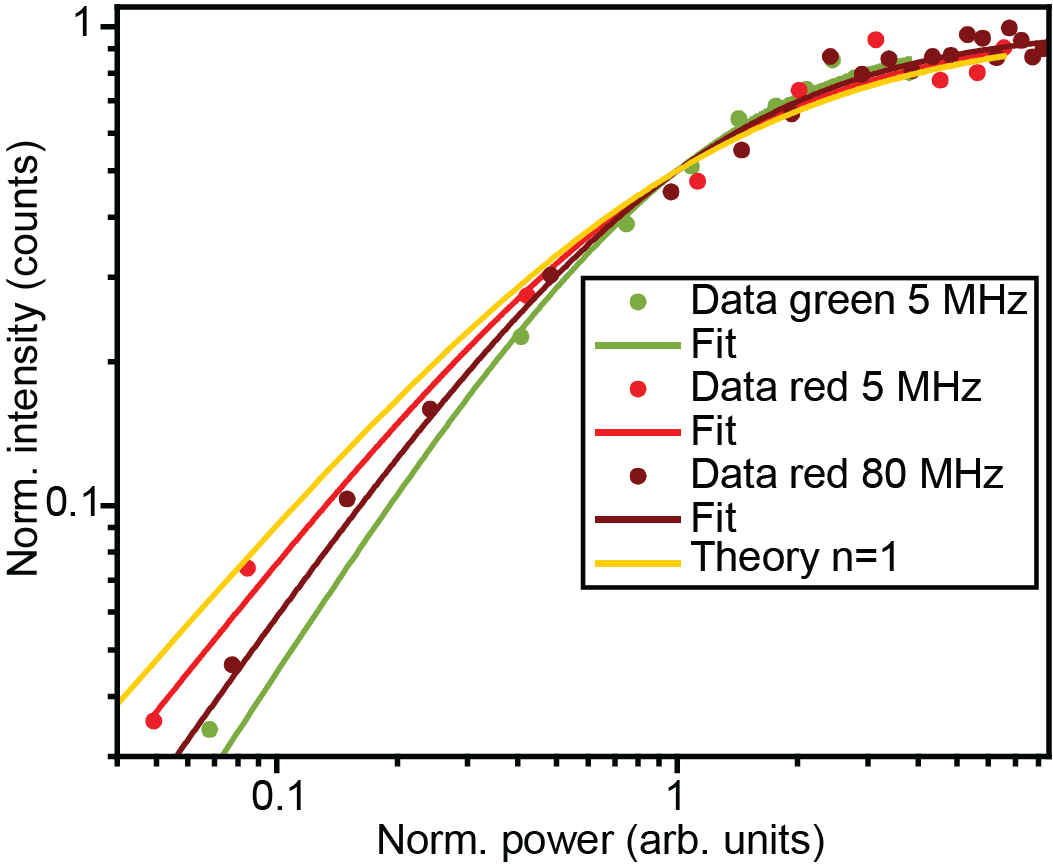}
	\caption{Peak area of emitter 1 as a function of excitation power in a double--logarithmic plot for red ($5$ and $80$ MHz) and green ($5$ MHz) excitation and the theoretical curve for an one--photon excitation process in yellow.}
	\label{suppfig:Power}
\end{figure}

A non--resonantly driven two--level system saturates with increasing excitation power. 
We investigated the behaviour for different excitation wavelengths, namely red $638~\si{\nm}$ with a repetition rate of $5$ and $80$ MHz, and green $532~\si{\nm}$ with a repetition rate of $5~\si{\MHz}$. 
In Fig.\,\ref{suppfig:Power} we show the peak areas as a function of excitation power in a double--logarithmic plot. 

All data sets are fitted with  $I(P)=I_{\infty}\frac{\left(\frac{P}{P_{sat}}\right)^n}{\left(\frac{P}{P_{sat}}\right)^n +1}$ \cite{schell_non-linear_2016} and weighted with $w_i=1/y_i$ to compensate for the fact of fewer data points at low excitation energies. 
To compare the data sets, they are normalized with $I_{\infty}$ and $P_{sat}$.
If the system is excited in a one--photon process, the saturation should follow the formula for $n=1$ (yellow line in Fig.\,\ref{suppfig:Power}), what is expected since the laser energy is higher than the one of the emitter.
The fits are yielding to $n=1.09\pm0.16$ ($n=1.21\pm0.08$) for red $5$~($80$)~MHz and $n=1.33\pm0.12$ for green $5$~MHz excitation.

\section{Analysis of second--order autocorrelation measurements under non--resonant excitation}

\begin{figure}
	\centering
	\includegraphics[width=0.5\columnwidth]{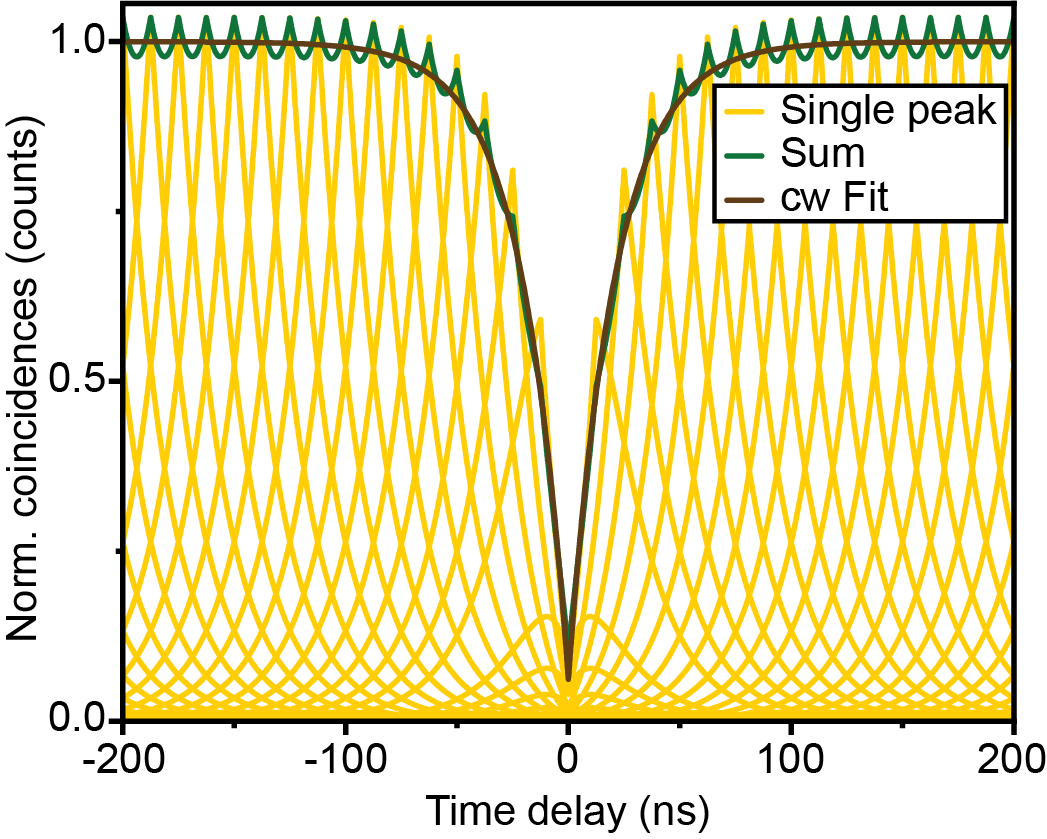}
	\caption{Simulation of a second--order correlation histogram for excitation with $80~\si{\MHz}$ and a emitter lifetime of $18.3~\si{\ns}$.}
	\label{suppfig:g2sim}
\end{figure}

The recorded time tag files were analyzed using ETA software~\cite{noauthor_extensible_2019} with a binning of $2048$~ps.
To explain why our pulsed $80$~\si{\MHz} second--order correlation measurement resembles a measurement under continuous--wave (cw) excitation, we simulated the resulting histogram, which we show in Fig.\,\ref{suppfig:g2sim}. 
Single peaks from each excitation pulse with a repetition rate of $80~\si{\MHz}$ with a lifetime of $18.3~\si{ns}$ are shown in yellow together with the resulting normalized histogram in green. 
The periodic modulations are not visible in our measurement in Fig.~$3$b and d due to noise. 
This strong overlap of the peaks justifies to treat the data like a cw measurement and fitted with $\text{g}^{(2)}(\tau)=\text{B}\left(1-\left(1-\text{g}^{(2)}(0)\right)\right)\exp(\tau/\tau_0)$ with the Poisson level B and the width $\tau_0$, as proofed with the fit of the summed simulation peaks in brown.

The extracted width of the dip in the non--resonantly excited second--order autocorrelation measurement of emitter 1 is $6.69\pm0.56$~ns, which deviates from the measured lifetime using lower repetition rates, and can be explained by the high excitation power. 
The second--order correlation function can be described by $\text{g}^{(2)}(\tau)=1-\left(1-\text{g}^{(2)}(0)\right)\exp(\tau/(\tau_l+1/ W_p))$~\cite{michler_quantum_2000} with the emitter lifetime $\tau_l$ and the pump rate into the excited state $W_p$. 
This leads to a narrowing of the dip for high excitation powers, which was the case in our measurement ($1.4~\si{\micro\watt} = 4.4~\text{P}_{\text{sat}}$).
This effect was not taken into account in the simplified simulation of the second--order autocorrelation measurement.

For the second--order autocorrelation measurement with $10$~\si{\MHz} repetition rate, the single peaks can be distinguished.
Here, the non--postselected second--order coherence g$^{(2)}(0)$ is given by the ratio of summed up coincidences in the center peak and average number of coincidences in the side peaks. 
Since the peaks still overlap in time, well--defined time windows to sum up the coincidences cannot be given, and we analyzed the data following the procedure described in~\cite{scholl_resonance_2019}.

\section{Methods for second--order autocorrelation measurements under resonant excitation}

\begin{figure}
	\centering
	\includegraphics[width=1\columnwidth]{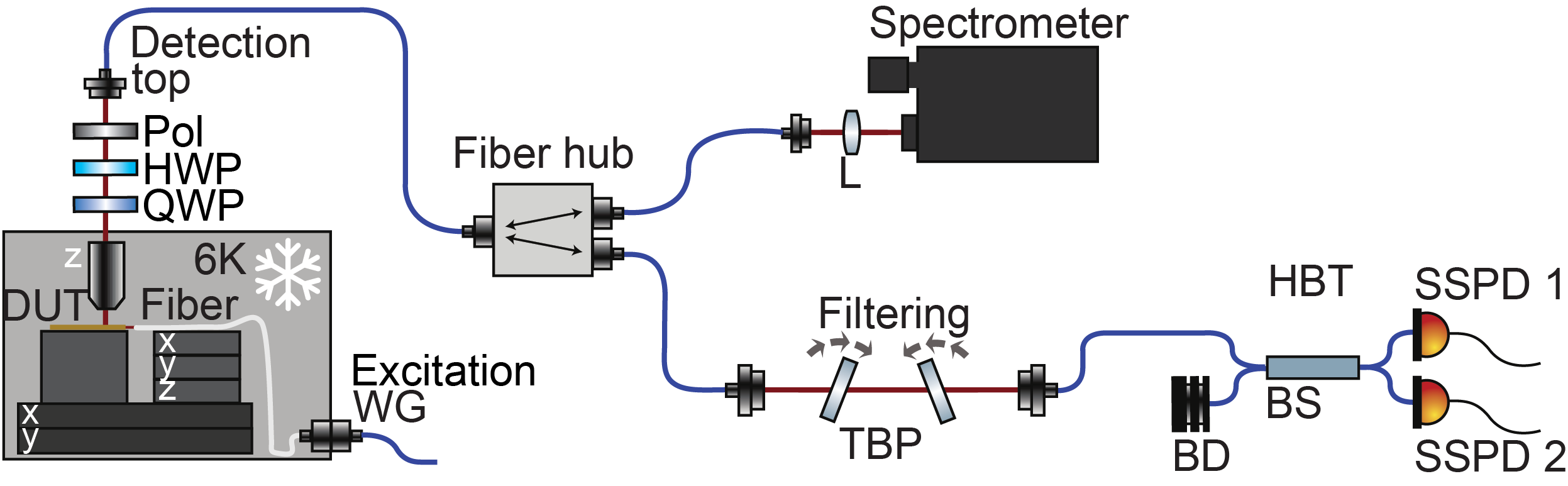}
	\caption{Side--excitation setup. The laser was coupled to the monolayer using a lensed fiber. The signal and scattered laser light was collected by the microscope objective, with the laser light being suppressed in a polarization suppression setup consisting of a quarter-- (QWP) and half--wave plate (HWP) and a linear polarizer (Pol). The fiber--coupled filtered signal was routed in the fiber hub onto the spectrometer or the Hanbury Brown and Twiss setup (HBT), which included a free--space filtering by two tunable bandpass filters (TBP). DUT device under test; BS beam splitter; L lens; BD beam dump; SSPD superconducting single photon detector.}
	\label{suppfig:setupRF}
\end{figure}

Figure \,\ref{suppfig:setupRF} shows the modified optical setup for resonant excitation of the quantum emitter. 
For this measurement, we excited the emitter from the side with a $50$~kHz linewidth continuous--wave diode laser by coupling it into the waveguide through a lensed fiber. 
The signal from the emitter as well as the remaining scattered laser light was collected by the microscope objective. 
To perform resonant excitation, the remaining laser has to be filtered with a polarization suppression setup. 
This requires well--defined excitation laser polarization, so that the remaining laser light is absorbed by a nanoparticle linear film polarizer after emitter excitation, and only a fraction of the signal of the quantum emitter, with its polarization perpendicular to that of the laser, is detected. 
In our setup, the laser was further spatially suppressed using a free--space to fiber coupling, with the core acting as a pinhole. 
The filtered fiber-coupled signal was sent either onto the CCD of a spectrometer or into a Hanbury Brown an Twiss setup to perform second--order autocorrelation measurements. 
This part of the setup remained the same as for previous measurements. 
In our setup, the excitation was not perfectly polarized, since the lensed fiber inside the cryostat is not polarization maintaining and the laser was scattering out of the waveguide via the sidewall roughness. 
Nevertheless, the quarter-- and halfwave plate (QWP, HWP) before the polarizer were aligned so that the remaining laser was minimized and on--the--fly optimization was possible during the measurements. 
The second--order autocorrelation measurement was recorded for a total amount of time of approx. $1.5$~h, and required $3$ realignments in addition to the on--the--fly polarization suppression optimization. 
The recorded timetag file was analyzed using readPTU~\cite{ballesteros_readptu_2019} with a time binning of $512$ ns. 
The resulting histogram was fitted using the formula given above for cw autocorrelation measurements.

\end{suppinfo}

\providecommand{\latin}[1]{#1}
\makeatletter
\providecommand{\doi}
  {\begingroup\let\do\@makeother\dospecials
  \catcode`\{=1 \catcode`\}=2 \doi@aux}
\providecommand{\doi@aux}[1]{\endgroup\texttt{#1}}
\makeatother
\providecommand*\mcitethebibliography{\thebibliography}
\csname @ifundefined\endcsname{endmcitethebibliography}
  {\let\endmcitethebibliography\endthebibliography}{}

\end{document}